\pdfoutput=1

\documentclass[11pt]{article}
\usepackage[]{acl}
\usepackage{times}
\usepackage{latexsym}
\usepackage[T1]{fontenc}
\usepackage[utf8]{inputenc}

\usepackage{microtype}
\usepackage{booktabs}
\usepackage{multirow}
\usepackage{multicol}
\usepackage{subcaption}
\usepackage[linesnumbered,ruled]{algorithm2e}
\usepackage{mathtools,amssymb,mathrsfs}
\usepackage{microtype}
\usepackage{enumitem}
\newcommand{\bb}[1]{\mathbf{#1}}

\newcommand{\bx}{\bb{x}}

\newcommand{\by}{\bb{y}}

\newcommand{\bz}{\bb{z}}

\newcommand{\datasetname}{PopBuTFy}

\title{Learning the Beauty in Songs: Neural Singing Voice Beautifier}

\author{Jinglin Liu \and Chengxi Li \and Yi Ren \and Zhiying Zhu \and Zhou Zhao \\
        \{jinglinliu,chengxili,rayeren,zhyingzh,zhaozhou\}@zju.edu.cn \\ Zhejiang University}

\begin{document}
\maketitle
\begin{abstract}
We are interested in a novel task, singing voice beautifying (SVB). Given the singing voice of an amateur singer, SVB aims to improve the intonation and vocal tone of the voice, while keeping the content and vocal timbre. Current automatic pitch correction techniques are immature, and most of them are restricted to intonation but ignore the overall aesthetic quality. Hence, we introduce Neural Singing Voice Beautifier (NSVB), the first generative model to solve the SVB task, which adopts a conditional variational autoencoder as the backbone and learns the latent representations of vocal tone. In NSVB, we propose a novel time-warping approach for pitch correction: Shape-Aware Dynamic Time Warping (SADTW), which ameliorates the robustness of existing time-warping approaches, to synchronize the amateur recording with the template pitch curve. Furthermore, we propose a latent-mapping algorithm in the latent space to convert the amateur vocal tone to the professional one. To achieve this, we also propose a new dataset containing parallel singing recordings of both amateur and professional versions. Extensive experiments on both Chinese and English songs demonstrate the effectiveness of our methods in terms of both objective and subjective metrics. Audio samples are available at~\url{https://neuralsvb.github.io}. Codes: \url{https://github.com/MoonInTheRiver/NeuralSVB}.

\end{abstract}

\section{Introduction}
\label{sec:intro}
The major successes of the artificial intelligent singing voice research are primarily in Singing Voice Synthesis (SVS)~\cite{lee2019adversarially,blaauw2020sequence,ren2020deepsinger,lu2020xiaoicesing,liu2021diffsinger} and Singing Voice Conversion (SVC)~\cite{sisman2020generative,li2021ppg,wang2021towardshighfidelity}. However, the Singing Voice Beautifying (SVB) remains an important and challenging endeavor for researchers. SVB aims to improve the intonation\footnote{Intonation refers to the accuracy of pitch in singing.} and the vocal tone of the voice, while keeping the content and vocal timbre\footnote{The differences between the vocal tone and vocal timbre is that: the former represents one's skills of singing, such as airflow controlling ability, muscle strength of vocal folds and vocal placement; the latter represents the identical, overall sound of one's vocal.}. SVB is extensively required both in the professional recording studios and the entertainment industries in our daily life, since it is impractical to record flawless singing audio.  

Nowadays in real-life scenarios, SVB is usually performed by professional sound engineers with adequate domain knowledge, who manipulate commercial vocal correction tools such as Melodyne\footnote{\url{https://www.celemony.com/en/start}} and Autotune\footnote{\url{https://www.antarestech.com/}} \cite{yong2018singing}. Most current automatic pitch correction works are shown to be an attractive alternative, but they may 1) show weak alignment accuracy~\cite{luo2018singing} or pitch accuracy~\cite{wager2020deep}; 2) cause the tuned recording and the reference recording to be homogeneous in singing style~\cite{yong2018singing}. Besides, they typically focus on the intonation but ignore the overall aesthetic quality (audio quality and vocal tone)~\cite{rosenzweig2021adaptive,zhuang2021karatuner}. 

To tackle these challenges, we introduce Neural Singing Voice Beautifier (NSVB), the first generative model to solve the SVB task, which adopts a Conditional Variational AutoEncoder (CVAE)~\cite{DBLP:journals/corr/KingmaW13,DBLP:conf/nips/SohnLY15} as the backbone to generate high-quality audio and learns the latent representation of vocal tone. In NSVB, we dichotomize the SVB task into pitch correction and vocal tone improvement: 1) To correct the intonation, a straightforward method is aligning the amateur recording with the template pitch curve, and then putting them together to resynthesize a new singing sample. Previous works~\cite{wada2017adaptive,luo2018singing} implemented this by figuring out the alignment through Dynamic Time Warping (DTW)~\cite{muller2007dynamic} or Canonical Time Warping (CTW)~\cite{NIPS2009_ctw}. We propose a novel Shape-Aware DTW algorithm, which ameliorates the robustness of existing time-warping approaches by considering the shape of the pitch curve rather than low-level features when calculating the optimal alignment path. 2) To improve the vocal tone, we propose a latent-mapping algorithm in the latent space, which converts the latent variables of the amateur vocal tone to those of the professional ones. This process is optimized by maximizing the log-likelihood of the converted latent variables. To retain the vocal timbre during the vocal tone mapping, we also propose a new dataset named  \datasetname~containing parallel singing recordings of both amateur and professional versions. 
Besides, thanks to the autoencoder structure, NSVB inherently supports semi-supervised learning, where the additional unpaired, unlabeled\footnote{``unpaired, unlabeled'' means the recordings sung by any people, in any vocal tone without label.} singing data could be leveraged to facilitate the learning of the latent representations.
Extensive experiments on both Chinese and English songs show that NSVB outperforms previous methods by a notable margin, and each component in NSVB is effective, in terms of both objective and subjective metrics.  The main contributions of this work are summarized as follows:
\begin{itemize}[leftmargin=*]
\item We propose the first generative model NSVB to solve the SVB task. NSVB not only corrects the pitch of amateur recordings, but also generates the audio with high audio quality and improved vocal tone, to which previous works typically pay little attention.

\item We propose Shape-Aware Dynamic Time Warping (SADTW) algorithm to synchronize the amateur recording with the template pitch curve, which ameliorates the robustness of the previous time-warping algorithm.

\item We propose a latent-mapping algorithm to convert the latent variable of the amateur vocal tone to the professional one's, and contribute a new dataset \datasetname to train the latent-mapping function. %

\item We design NSVB as a CVAE model, which supports the semi-supervised learning to leverage unpaired, unlabeled singing data for better performance.

\end{itemize}

\section{Related Works}
\subsection{Singing Voice Conversion}
Singing Voice Conversion (SVC) is a sub-task of Voice Conversion (VC)~\cite{berg2015gpu,serra2019blow,popov2021diffusion,liu2021diffsvc}, which transforms the vocal timbre (or singer identity) of one singer to that of another singer, while preserving the linguistic content and pitch/melody information~\cite{li2021ppg}. Mainstream SVC models can be grouped into three categories~\cite{zhao2020voice}: 1) parallel spectral feature mapping models, which learn the conversion function between source and target singers relying on parallel singing data~\cite{villavicencio2010applying,kobayashi2015statistical,sisman2019singan}; 2) Cycle-consistent Generative Adversarial Networks (CycleGAN)~\cite{zhu2017unpaired,kaneko2019cyclegan}, where an adversarial loss and a cycle-consistency loss are concurrently used to learn the forward and inverse mappings simultaneously~\cite{sisman2020generative}; 3) encoder-decoder models, such as PPG-SVC~\cite{li2021ppg}, which leverage a singing voice synthesis (SVS) system for SVC~\cite{DBLP:conf/interspeech/ZhangYLWZWXLY20}, and auto-encoder~\cite{qian2019autovc,wang2021vqmivc,yuan2020improving} based SVC~\cite{wang2021towardshighfidelity}. The models of the latter two categories can be utilized with non-parallel data. In our work, we aim to convert the intonation and the vocal tone while keeping the content and the vocal timbre, which is quite different from the SVC task. %

\begin{figure*}[!h]
    \centering
        \includegraphics[width=0.95\textwidth]{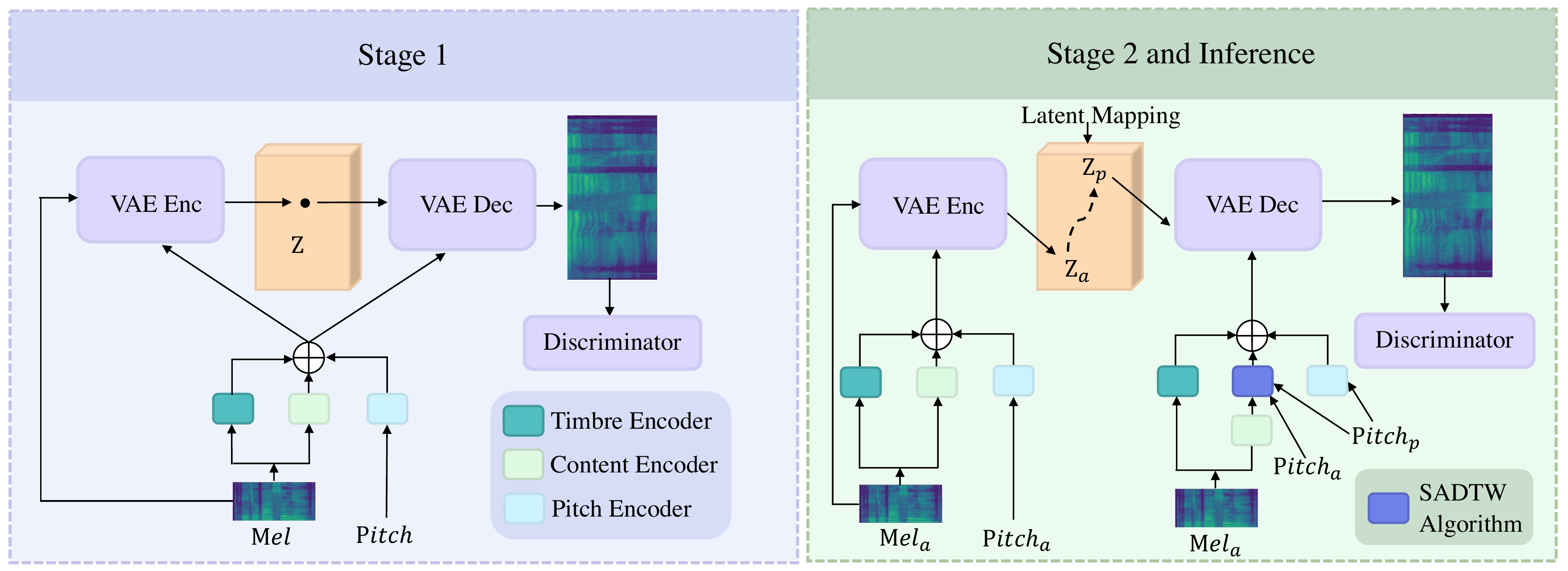}
        \caption{The overview of NVSB. The training process consists of 2 stages, and the second stage shares the same pipeline with the inference stage. ``VAE Enc'' means the encoder of CVAE; ``VAE Dec'' means the decoder of CVAE; ``$\text{Mel}$'' means the mel-spectrogram; ``$\bz$'' means the latent variable of the vocal tone; the ``$a$''/``$p$'' subscript means the amateur/professional version.}
        \label{fig:method_pipeline}
\end{figure*}

\subsection{Automatic Pitch Correction}
Automatic Pitch Correction (APC) works attempt to minimize the manual effort in modifying the flawed singing voice~\cite{yong2018singing}. \citet{luo2018singing} propose Canonical Time Warping (CTW)~\cite{NIPS2009_ctw,zhou2012generalized} which aligns amateur singing recordings to professional ones according to the pitch curves only. \citet{wager2020deep} propose a data-driven approach to predict pitch shifts depending on both amateur recording and its accompaniment. \citet{rosenzweig2021adaptive} propose a pitch shift method for Cappella recordings. \citet{zhuang2021karatuner} propose a pitch-controllable SVS system to resynthesize the audio with correctly predicted pitch curves. Besides modifying pitch, \citet{yong2018singing} propose to modify pitch and energy information to improve the singing expressions of an amateur singing recording. However, this method heavily relies on a reference recording, causing the tuned recording and the reference recording to be homogeneous in singing style~\cite{zhuang2021karatuner}. Our work adopts the non-parametric and data-free pitch correction method like \citet{luo2018singing}, but improves the accuracy of alignment.

\section{Methdology}
In this section, we describe the overview of NSVB, which is shown in Figure~\ref{fig:method_pipeline}. At Stage 1 in the figure, we reconstruct the input mel-spectrogram through the CVAE backbone (Section~\ref{sec:backbonemodel}) based on the pitch, content and vocal timbre conditions extracted from the input by the pitch encoder, content encoder and timbre encoder, and optimize the CVAE by maximizing evidence lower bound and adversarial learning. At Stage 2/Inference in the figure, firstly we infer the latent variable $\bz_a$ based on the amateur conditions; secondly we prepare the amateur content vectors aligned with the professional pitch by SADTW algorithm (Section~\ref{sec:sadtw}); thirdly we map $\bz_a$ to $\bz_p$ by the latent-mapping algorithm (Section~\ref{sec:latentmapping}); finally, we mix the professional pitch, the aligned amateur content vectors, and the amateur vocal timbre to obtain a new condition, which is leveraged along with the mapped $\bz_p$ by the decoder of CVAE to generate a new beautified mel-spectrogram. The training/inference details and model structure of each component in NSVB are described in Section~\ref{sec:traininginference} and Section~\ref{sec:model_structure}.

\subsection{Conditional Variational Generator with Adversarial Learning}
\label{sec:backbonemodel}
\begin{figure}[!htb]
	\centering
	\includegraphics[width=0.4\textwidth]{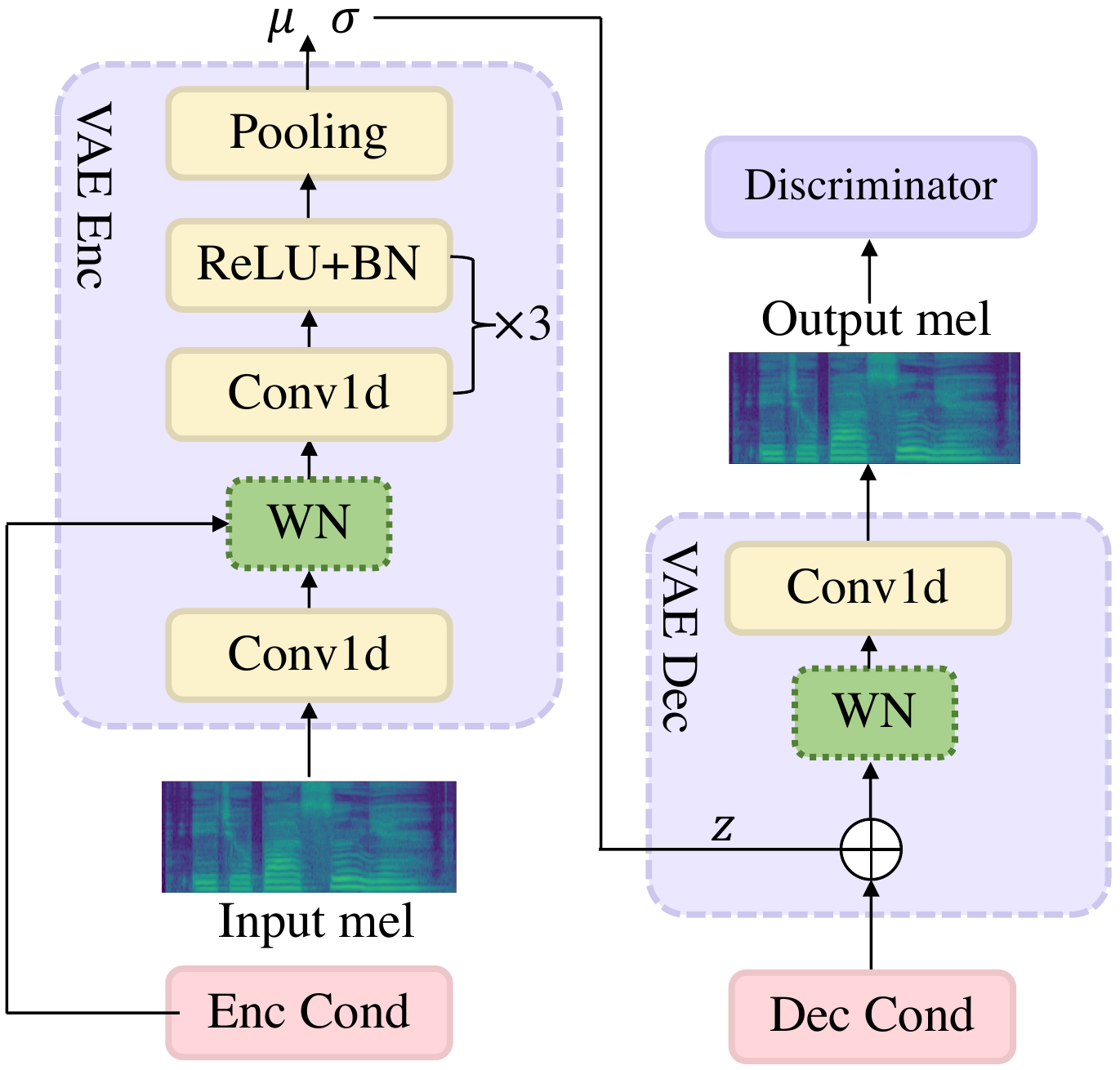}
	\caption{The CVAE backbone in NSVB. ``Enc/Dec Cond'' means the conditions for the encoder/decoder; ``Conv1d'' means the 1-D convolutional layer; ``Pooling'' means the average pooling layer; $\mu$ and $\sigma$ represent the approximated mean and log scale standard deviation parameters in the posterior Gaussian distribution; $\bz$ is the sampled latent variable. }
	\label{fig:method_cvae}  
\end{figure}

As shown in Figure~\ref{fig:method_cvae}, to generate audio with high quality and learn the latent representations of vocal tone, we introduce a Conditional Variational AutoEncoder (CVAE)~\cite{DBLP:journals/corr/KingmaW13,DBLP:conf/nips/SohnLY15} as the mel-spectrogram generator, with the optimizing objective of maximizing the evidence lower bound (ELBO) of the intractable marginal log-likelihood of mel-spectrogram $\log p_{\theta}(\bx|c)$:
\begin{align*}
& \log p_{\theta}(\bx|c) \ge \mathbf{ELBO}(\phi, \theta) \equiv \\ 
& E_{\bz \sim q_{\phi}(\bz|\bx,c)} \left[\log p_{\theta}(\bx|\bz, c)- \log \frac{q_{\phi}(\bz|\bx, c)}{p(\bz)}\right] \nonumber ,
\end{align*}
where $\bx$, $c$, $\bz$ denote the input/output mel-spectrogram, the mix of content, vocal timbre and pitch conditions, and the latent variable representing the vocal tone respectively; $\phi$ and $\theta$ denote the model parameters of CVAE encoder and CVAE decoder; $q_{\phi}(\bz|\bx,c)$ is the posterior distribution approximated by the CVAE encoder; $p_{\theta}(\bx|\bz, c)$ is the likelihood function that generates mel-spectrograms given latent variable $\bz$ and condition $c$; $p(\bz)$ is the prior distribution of the latent variables $\bz$, and we choose the standard normal distribution as $p(\bz)$ for simplification. Furthermore, to address the over-smoothing problem~\cite{2019autovc} in CVAE, we utilize an adversarial discriminator ($\mathcal{D}$)~\cite{mao2017least} to refine the output mel-spectrogram:
\begin{align}
    & L_{adv} (\phi, \theta) = \mathbb{E} [(\mathcal{D}(\widetilde{\bx})-1)^{2}], \nonumber \\
   \label{eq:adv_d}
    & L_{adv} (\mathcal{D}) = \mathbb{E} [(\mathcal{D}(\bx)-1)^{2}]+\mathbb{E} [\mathcal{D}(\widetilde{\bx})^{2}], 
\end{align}
where $\bx$ is the ground-truth and $\widetilde{\bx}$ is the output of CVAE.
The descriptions for the model structure of each component are in Section~\ref{sec:model_structure}.

\subsection{Shape-Aware Dynamic Time Warping}
\label{sec:sadtw}
To implement the pitch correction, a straightforward method is aligning the amateur recording with the template pitch curve, and then concatenating them to resynthesize a new singing sample with improved intonation. Since the source pitch curve of amateur recordings and template one show a high degree of natural correlation along the time axis, applying a proper time-warping algorithm on them is crucial. However, original DTW~\cite{muller2007dynamic} could result in a poor alignment when certain parts of the axis move to higher frequencies, and other parts to lower ones, or vice versa~\cite{sundermann2003vtln}. \citet{luo2018singing} adopt an advanced algorithm CTW~\cite{NIPS2009_ctw}, which combines the canonical correlation analysis (CCA) and DTW to extract the feature sequences of two pitch curves, and then apply DTW on them. However, the alignment accuracy of CTW leaves much to be desired. 

\begin{figure}[!htb]
	\centering
	\includegraphics[width=0.4\textwidth]{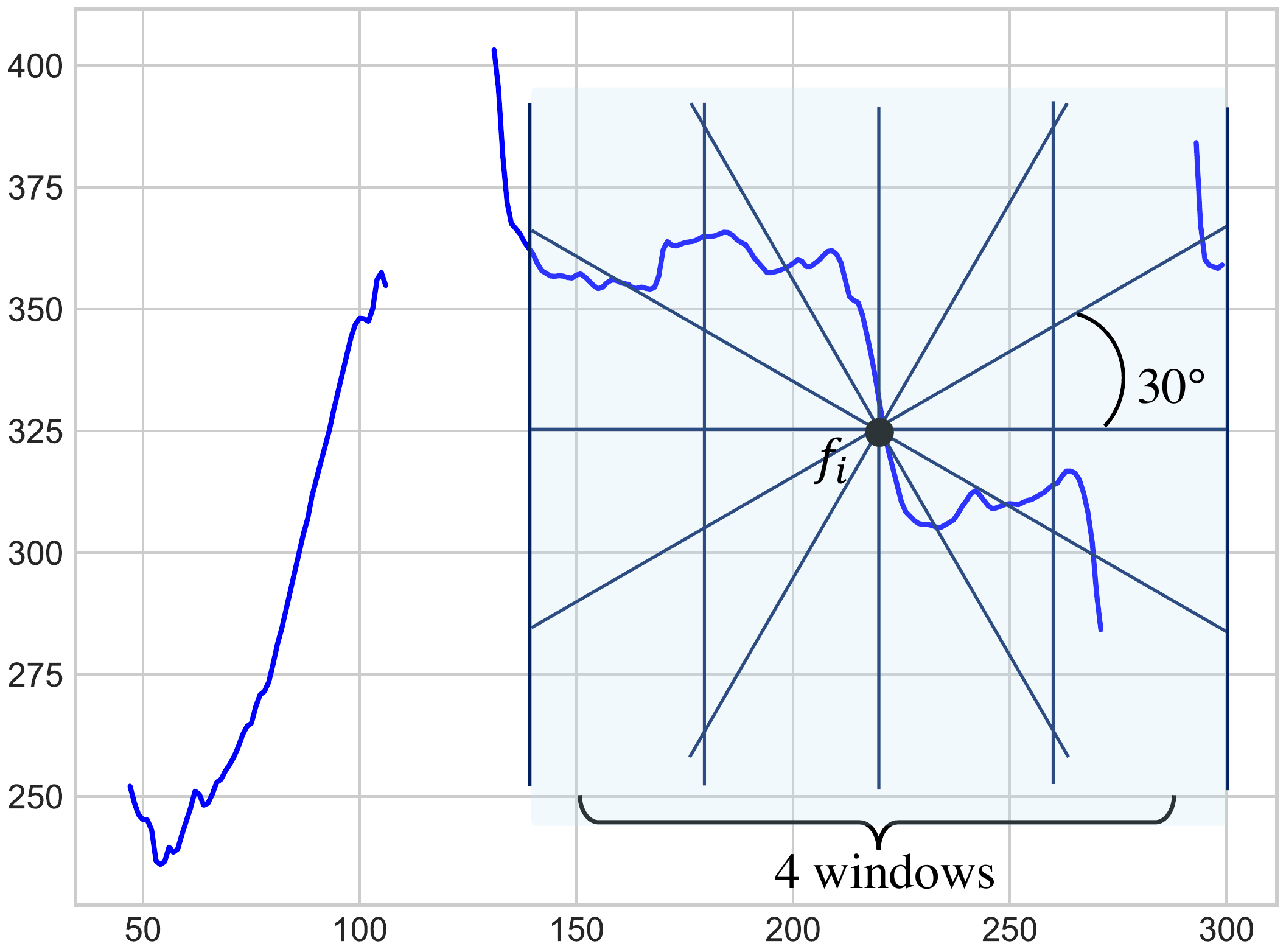}
	\caption{The shape descriptor in SADTW. The blue curve represents pitch; the horizontal axis means time; the vertical axis means F0-frequency. There are $m=4$ windows, $n=6$ angles to divide neighbor points of $f_i$.}
	\label{fig:method_sadtw}
\end{figure}

We elaborate a non-parametric and data-free algorithm, Shape-Aware DTW (SADTW), based on the prior knowledge that the source pitch curve and the template one have analogous local shape contours. Specifically, we replace the Euclidean distance in the original DTW distance matrix with the shape context descriptor distance. The shape context descriptor of a time point $f_i$ in one pitch curve is illustrated in Figure~\ref{fig:method_sadtw}. Inspired by ~\cite{mori2005efficient}, we divide the data points around $f_i$ into $m * n$ bins by $m$ time windows and $n$ angles. We calculate the number of all points falling in the $k$-th bin. Then the descriptor for $f_i$ is defined as the histogram $h_i \in \mathcal{R}^{m * n}$:
\begin{align*}
    h_i(k) = | \{f_j \neq f_i, f_j \in bin(k) \} | ,
\end{align*}
where $|\cdot|$ means the cardinality of a set. This histogram represents the distribution over relative positions, which is a robust, compact and discriminative descriptor. Then, it is natural to use the $\mathcal{X}^2$-test statistic on this distribution descriptor as the ``distance'' of two points $f_a$ and $f_p$:
\begin{align*}
    C (a,p) = \frac{1}{2} \sum\limits_{k=1}^{m*n} \frac{\left[ h_a(k)-h_p(k) \right] ^2}{h_a(k)+h_p(k)} ,
\end{align*}
where $h_a$ and $h_p$ are the normalized histograms corresponding to the point $f_a$ from the amateur pitch curve and the point $f_p$ from the template pitch curve. $C (a,p)$ ranges from 0 to 1. Finally, we run DTW on the distance matrix $C$ to obtain the alignment with least distance cost between two curves. 

\subsection{Latent-mapping Algorithm}
\label{sec:latentmapping}
Define a pair of mel-spectrograms $(\bx_a, \bx_p)$: the contents of $\bx_a$ and $\by_p$ are the same sentence of a song from the same singer\footnote{The singers all major in vocal music.}, who sings these two recordings using the amateur tone and the professional tone respectively. Given the CVAE model, we can infer the posterior distribution $q_{\phi}(\bz_a|\bx_a,c_a)$ and $q_{\phi}(\bz_p|\bx_p,c_p)$ corresponding to $\bx_a$ and $\bx_p$ through the encoder of CVAE. To achieve the conversion of vocal tone, we introduce a mapping function $\mathcal{M}$ to convert the latent variables from $q_{\phi}(\bz_a|\bx_a,c_a)$ to $q_{\phi}(\bz_p|\bx_p,c_p)$. Concretely, we sample a latent variable of amateur vocal tone $\bz_a$ from $q_{\phi}(\bz_a|\bx_a,c_a)$, and map $\bz_a$ to $\mathcal{M}(\bz_a)$. Then, $\mathcal{M}$ can be optimized by minimizing the negative log-likelihood of $\mathcal{M}(\bz_a)$:
\begin{align*}
    L_{map1}(\mathcal{M}) = - \log q_{\phi}(\mathcal{M}(\bz_a)|\bx_p,c_p).
\end{align*}

Define $\hat{c}_p$ as the mix of 1) the content vectors from the amateur recording aligned by SADTW, 2) vocal timbre embedding encoded by timbre encoder, and 3) template pitch\footnote{During training, template pitch is extracted from the waveform corresponding to $\bx_p$.} embeddings encoded by pitch encoder. To make sure the converted latent variable could work well together with $\hat{c}_p$ to generate a high-quality audio sample (with the correct pitch and improved vocal tone), we send $\mathcal{M}(\bz_a)$ to the CVAE decoder to generate $\hat{\bx}$, and propose an additional loss:
\begin{equation*}
    L_{map2} (\mathcal{M}) = \| \hat{\bx} - \bx_p \|_1 + \lambda (\mathcal{D}(\hat{\bx})-1)^{2},
\end{equation*}
where $\mathcal{D}$ has been optimized by Eq.~\eqref{eq:adv_d}; $\lambda$ is a hyper-parameter.

\subsection{Training and Inference}
\label{sec:traininginference}
There are two training stages for NSVB: in the first training stage, we optimize CVAE by minimizing the following loss function
\begin{equation*} 
    L(\phi, \theta) = - \mathbf{ELBO}(\phi, \theta) + \lambda L_{adv}(\phi, \theta),
\end{equation*}
and optimize the discriminator ($\mathcal{D}$) by minimizing Eq.~\eqref{eq:adv_d}. Note that, the first stage is the reconstruction process of mel-spectrograms, where any unpaired, unlabeled singing data beyond \datasetname~could be leveraged to facilitate the learning of the latent representations. In the second training stage, we optimize $\mathcal{M}$ on the parallel dataset \datasetname~by minimizing the following loss function
\begin{equation*} 
    L(\mathcal{M}) = L_{map1}(\mathcal{M}) + L_{map2}(\mathcal{M}).
\end{equation*}
$\phi$, $\theta$, and $\mathcal{D}$ are not optimized in this stage.

In inference, the encoder of CVAE encodes $\bx_a$ with the condition $c_a$ to predict $\bz_a$. Secondly, we map $\bz_a$ to $\mathcal{M}(\bz_a)$, and run SADTW to align the amateur recordings with the template pitch curve. The template pitch curve can be derived from a reference recording with good intonation or a pitch predictor with the input of music notes. Then, we obtain $\hat{c}_p$ defined in Section~\ref{sec:latentmapping} and send $\mathcal{M}(\bz_a)$ together with $\hat{c}_p$ in the decoder of CVAE to generate $\hat{\bx}$. Finally, by running a pre-trained vocoder conditioned on $\hat{\bx}$, a new beautified recording is produced.
\subsection{Model Structure}
\label{sec:model_structure}
The encoder of CVAE consists of a 1-D convolutional layer (stride=4), an 8-layer WaveNet structure~\cite{vanwavenet,rethage2018wavenet} and 3 1-D convolutional layers (stride=2) with ReLU activation function and batch normalization followed by a mean pooling, which outputs the mean and log scale standard deviation parameters in the posterior distribution of $\bz$. The decoder of CVAE consists of a 4-layer WaveNet structure and a 1-D convolutional layer, which outputs the mel-spectrogram with 80 channels. The discriminator adopts the same structure as \cite{wu2020adversarially}, which consists of multiple random window discriminators. The latent-mapping function is composed of 2 linear layers to encode the vocal timbre as the mapping condition, and 3 linear layers to map $\bz_a$. The pitch encoder is composed of 3 convolutional layers. In addition, given a singing recording, 1) to obtain its content vectors, we train an Automatic Speech Recognition (ASR) model based on Conformer~\cite{DBLP:conf/interspeech/GulatiQCPZYHWZW20} with both speech and singing data, and extract the hidden states from the ASR encoder (viewed as the content encoder) output as the linguistic content information, which are also called phonetic posterior-grams (PPG); 2) to obtain the vocal timbre, we leverage the open-source API resemblyzer\footnote{\url{https://github.com/resemble-ai/Resemblyzer}} as the timbre encoder, which is a deep learning model designed for speaker verification~\cite{wan2018generalized}, to extract the identity information of a singer. More details of model structure can be found in Appendix \ref{appendxi:detailed_model}.

\section{Experiments}
\subsection{Experimental Setup}
\label{section:setup}
In this section, we first introduce \datasetname, the dataset for SVB, and then describe the implementation details in our work. Finally, we explain the evaluation method we adopt in this paper.
\paragraph{Dataset}
Since there is no publicly available high-quality, unaccompanied and parallel singing dataset for the SVB task, we collect and annotate a dataset containing both Chinese Mandarin and English pop songs: \datasetname. To collect \datasetname~for SVB, the qualified singers majoring in vocal music are asked to sing a song twice, using the amateur vocal tone for one time and the professional vocal tone for another. Note that some of the amateur recordings are sung off-key by one or more semi-tones for the pitch correction sub-task. The parallel setting could make sure that the personal vocal timbre will keep still during the beautifying process. In all, \datasetname~consists of 99 Chinese pop songs ($\sim$10.4 hours in total) from 12 singers and 443 English pop songs ($\sim$40.4 hours in total) from 22 singers. All the audio files are recorded in a professional recording studio by qualified singers, male and female. Every song is sampled at 22050 Hz with 16-bit quantization. 
We randomly choose 274 pieces in Chinese and 617 pieces in English for validation and test.
For subjective evaluations, we choose 60 samples in the test set from different singers, half in Chinese and English. All testing samples are included for objective evaluations. 
\paragraph{Implementation Details}
We train the Neural Singing Beautifier on a single 32G Nividia V100 GPU with the batch size of 64 sentences for both 100k steps in Stage 1 and Stage 2 respectively. Besides \datasetname, we pre-train the ASR model (used for PPG extraction) leveraging the extra speech datasets: AISHELL-3~\cite{AISHELL-3_2020} for Chinese and LibriTTS~\cite{DBLP:conf/interspeech/ZenDCZWJCW19} for English. For the semi-supervised learning mentioned in Section~\ref{sec:intro} and Section~\ref{sec:traininginference}, we leverage an internal Chinese singing dataset ($\sim$30 hours without labeled vocal tone) in the first training stage described in Section~\ref{sec:traininginference} for Chinese experiments. The output mel-spectrograms of our model are transformed into audio samples using a HiFi-GAN vocoder~\cite{NEURIPS2020_hifigan} trained with singing data in advance. We set the $\lambda$ metioned in Section~3.3 to $0.1$. We transform the raw waveform with the sampling rate 22050 Hz into mel-spectrograms with the frame size 1024 and the hop size 128. We extract $F_0$ (fundamental frequency) as pitch information from the raw waveform using Parselmouth\footnote{\url{https://github.com/YannickJadoul/Parselmouth}}, following \citet{wu2020adversarially,blaauw2020sequence,ren2020deepsinger}. To obtain the ground truth pitch alignment between the amateur recordings and the professional ones for evaluating the accuracy of pitch alignment algorithm, we run the Montreal Forced Aligner tool~\cite{mcauliffe2017montreal} on all the singing recordings to obtain their alignments to lyrics. Then the ground-truth pitch alignment can be derived since the lyrics are shared in a pair of data in \datasetname. %
\paragraph{Performance Evaluation}
We employ both subjective metrics: Mean Opinion Score (MOS), Comparison Mean Opinion Score (CMOS), and an objective metric: Mean Cepstral Distortion (MCD) to evaluate the audio quality on the test-set. Besides, we use F0 Root Mean Square Error (F0 RMSE) and Pitch Alignment Accuracy (PAA) to estimate the pitch correction performance. For audio, we analyze the MOS and CMOS in two aspects: audio quality (naturalness, pronunciation and sound quality) and vocal tone quality. MOS-Q/CMOS-Q and MOS-V/CMOS-V correspond to the MOS/CMOS of audio quality and vocal tone quality respectively. More details about subjective evaluations are placed in Appendix \ref{appendxi:detailed_sub}.

\subsection{Main Results}

In this section, we conduct extensive experiments to present our proposed model in regard to 1) the performance of pitch conversion; 2) the audio quality and vocal tone quality.
\subsubsection{Pitch Correction}
\begin{figure*}[!t]
    \centering
	\includegraphics[width=\textwidth]{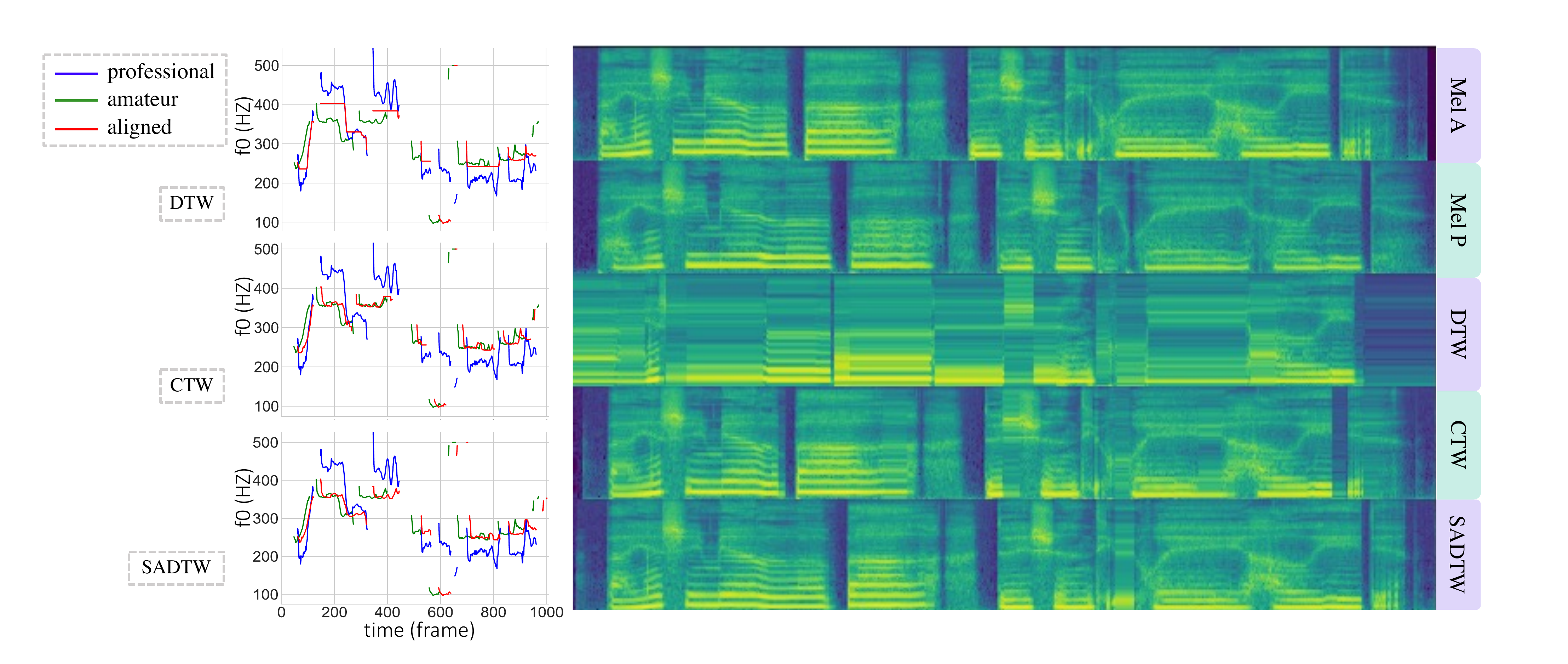}
	
	\caption{The behavior of DTW, CTW and SADTW. 1) In the left panel of the figure, we align the pitch curve of the amateur recording to the professional one's. It can be seen that DTW perform terribly; CTW fails at many parts; SADTW perform well as expectation. 2) In the right panel of the figure, we use the alignments obtained from these time-warping algorithm on pitch curves to align the amateur mel-spectrogram to the professional one. It shows that only SADTW could provide an alignment which preserves the content information in the amateur recording well and make the aligned result match the professional recording along the time axis.}
    \label{fig:alignment_case_vis}
\end{figure*}

Firstly, we provide the comparison among time-warping algorithms in terms of PAA in Table~\ref{tab:pitch_align_acc}. \textit{Normed DTW} means two pitch curves will be normalized before running \textit{DTW}~\cite{muller2007dynamic}; \textit{CTW} means the Canonical Time Warping~\cite{NIPS2009_ctw}, which is used for pitch correction in \citet{luo2018singing}. It can be seen that, \textit{SADTW} surpasses existing methods by a large margin. We also visualize an alignment example of \textit{DTW}, \textit{CTW}, and \textit{SADTW} in Figure~\ref{fig:alignment_case_vis}.

\newcommand{\plotpitch}[2]{
    \begin{subfigure}{0.3\textwidth}
	\centering
	\includegraphics[width=\textwidth, clip=true]{#1}
	\vspace{-4mm}
    \end{subfigure}
}
\newcommand{\plotmel}[2]{
    \begin{subfigure}{0.3\textwidth}
	\centering
	\includegraphics[width=\textwidth, clip=true]{#1}
	\vspace{-2mm}
	\caption{\textit{#2}}
	\vspace{2mm}
    \end{subfigure}
}

Secondly, to check whether the amateur recordings are corrected to the good intonation after being beautified by NSVB, we calculate the F0 RMSE metric of the amateur recordings and the audio generated by NSVB, and list the results in Table~\ref{tab:correction_res}. We can see that F0 RMSE has been improved significantly, which means NSVB successfully achieve pitch correction. 
\begin{table}[!htb]
    \small
    \centering
    \caption{The Pitch Alignment Accuracy of different algorithms on Chinese and English songs.}
    \begin{tabular}{ l | c | c }
    \toprule
    \multirow{2}*{Algorithm} &  \multicolumn{2}{c}{PAA (\%)}\\
    \cline{2-3}
           & Chinese & English \\
    \midrule
    \textit{DTW} & 66.94 & 63.90 \\
    \textit{Normed DT}W & 65.19 & 62.86\\
    \textit{CTW}  & 71.35 & 69.28\\
    \textit{SADTW}  & \textbf{79.45} & \textbf{78.64}\\
    \bottomrule
    \end{tabular}

    \label{tab:pitch_align_acc}
\end{table}
\begin{table}[!htb]
    \small
    \centering
    \caption{The F0 RMSE of the original amateur audio and the beautified audio on Chinese and English datasets. ``GT Amateur'' means the ground-truth amateur recordings.}
    \begin{tabular}{ l | c | c }
    \toprule
    \multirow{2}*{Algorithm} &  \multicolumn{2}{c}{F0 RMSE (Hz)}\\
    \cline{2-3}
           & Chinese & English  \\
    \midrule
    \textit{GT Amateur} & 25.11 & 23.75\\
    \textit{NVSB} & \textbf{6.96} & \textbf{7.29}\\
    \bottomrule
    \end{tabular}
    
    \label{tab:correction_res}
\end{table}

\subsubsection{Audio Quality and Vocal Tone Quality}

To thoroughly evaluate our proposed model in audio quality and vocal tone quality, we compare subjective metric MOS-Q, MOS-V and objective metric MCD of audio samples generated by NVSB with the systems including: 1) \textit{GT Mel}, amateur~(A) and professional~(P) version, where we first convert ground truth audio into mel-spectrograms, and then convert the mel-spectrograms back to audio using HiFi-GAN introduced in Section~\ref{section:setup}; 2) \textit{Baseline}: the baseline model for SVB based on WaveNet with the number of parameters similar to \textit{NSVB}, which adopts the same pitch correction method (SADTW) as \textit{NSVB} does, and takes in the condition $\hat{c}_p$ defined in Section~\ref{sec:latentmapping} to generate the mel-spectrogram optimized by the $L_1$ distance to $x_p$. MCD is calculated using the audio samples of \textit{GT Mel P} as references. 

The subjective and objective results on both Chinese and English datasets are shown in Table \ref{tab:main_sub}. We can see that 1) \textit{NSVB} achieves promising results, with MOS-Q being less than those for ground truth professional recordings by only 0.1 and 0.12 on both datasets; 2) \textit{NSVB} surpasses the \textit{GT Mel A} in terms of MOS-V by a large margin, which indicates that \textit{NSVB} successfully accomplishes the vocal tone improvement. 3) \textit{NSVB} surpasses the baseline model on all the metrics distinctly, which proves the superiority of our proposed model; 4) \textit{GT Mel P}, \textit{NSVB} and \textit{Baseline} all outperform \textit{GT Mel A} in terms of MOS-V, which demonstrates that the proposed dataset \datasetname~is reasonably labeled in respect of vocal tone. %

\begin{table}[!htb]
    \small
    \centering
    \caption{The Mean Opinion Score in audio quality (MOS-Q), vocal tone (MOS-V) with 95\% confidence intervals and the Mean Cepstral Distortion (MCD) comparisons with ground-truth singing recordings and baseline model.}
    \begin{tabular}{ l | c | c | c}
    \toprule
    Method &  MOS-Q & MOS-V & MCD\\
    \midrule
    \multicolumn{3}{l}{\textbf{Chinese}} \\
    \midrule
    \textit{GT Mel P} & 4.21 $\pm$ 0.06 & 4.27 $\pm$ 0.10 & -\\
    \textit{GT Mel A} & 4.11 $\pm$ 0.07 & 3.51 $\pm$ 0.13 & -\\
    \midrule
    \textit{Baseline} & 3.90 $\pm$ 0.09& 3.58 $\pm$ 0.18 & 7.609\\
    \textit{NVSB} & \textbf{4.11 $\pm$ 0.07} & \textbf{3.69 $\pm$ 0.17} & \textbf{7.068} \\
    \midrule
    \multicolumn{3}{l}{\textbf{English}} \\
    \midrule
    \textit{GT Mel P} & 3.96 $\pm$ 0.11 & 3.96 $\pm$ 0.18 & -\\
    \textit{GT Mel A} & 3.67 $\pm$ 0.11 & 3.36 $\pm$ 0.19 & -\\
    \midrule
    \textit{Baseline} &  3.65 $\pm$ 0.12 & 3.37 $\pm$ 0.19 & 8.166 \\
    \textit{NVSB} & \textbf{3.84 $\pm$ 0.06} & \textbf{3.63 $\pm$ 0.18} &\textbf{7.992}\\
    \bottomrule
    \end{tabular}

    \label{tab:main_sub}
\end{table}

\subsection{Ablation Studies}

We conduct some ablation studies to demonstrate the effectiveness of our proposed methods and some designs in our model, including latent-mapping, additional loss $\mathcal{L}_{map2}$ in the second training stage, and semi-supervised learning with extra unpaired, unlabeled data on Chinese songs. 

\subsubsection{Latent Mapping}

We compare audio samples from NSVB with and without latent-mapping in terms of CMOS-V and MCD. From Table \ref{tab:ab1}, we can see that the latent-mapping brings CMOS-V and MCD gains, which demonstrates the improvements in vocal tone from latent-mapping in our model. We visualize linear-spectrograms of \textit{GT Mel A}, \textit{GT Mel P}, \textit{NSVB}, \textit{NSVB w/o mapping} in Appendix \ref{appendxi:visualize}. The patterns of high-frequency parts in \textit{NVSB} samples are comparatively similar to those in \textit{GT Mel P} samples while \textit{NSVB w/o mapping} sample resembles \textit{GT Mel A} samples.
\begin{table}[!htb]
    \small
    \centering
    \caption{The Comparison Mean Opinion Score in vocal tone (CMOS-V) and the Mean Ceptral Distortion (MCD) results of singing audio samples for latent mapping. 
    }
    \begin{tabular}{ l | c | c }
    \toprule
    Method & CMOS-V & MCD\\
    \midrule
    \multicolumn{2}{l}{\textbf{Chinese}} \\
    \midrule
    \textit{NVSB}  & 0.000 & \textbf{7.068}\\
    \textit{NVSB w/o mapping}  & -0.100 & 7.069\\  %
    \midrule
    \multicolumn{2}{l}{\textbf{English}} \\
    \midrule
    \textit{NVSB}  & 0.000 & \textbf{7.992}\\
    \textit{NVSB w/o mapping}  &  -0.330 & 8.115\\ %
    \bottomrule
    \end{tabular}

    \label{tab:ab1}
\end{table}

\subsubsection{Additional Loss $\mathcal{L}_{map2}$}
As shown in Table~\ref{tab:ab2}, all the compared metrics show the effectiveness of $\mathcal{L}_{map2}$, which means that the additional loss $\mathcal{L}_{map2}$ is beneficial to optimizing the latent mapping function $\mathcal{M}$, working as a complement to the basic loss $\mathcal{L}_{map1}$.
\begin{table}[!htb]
    \small
    \centering
    \caption{The Comparison Mean Opinion Score in audio quality (CMOS-Q), vocal tone (CMOS-V) and the Mean Ceptral Distortion (MCD) of singing audio samples.}
    \begin{tabular}{ l | c| c | c }
    \toprule
    Method & CMOS-Q & CMOS-V & MCD \\
    \midrule
    \multicolumn{2}{l}{\textbf{Chinese}} \\
    \midrule
    \textit{NVSB} & 0.000 & 0.000 & \textbf{7.068} \\
    \textit{NVSB w/o $\mathcal{L}_{map2}$} &-0.213 & -0.760 & 7.237 \\ %
    \midrule
    \multicolumn{2}{l}{\textbf{English}} \\
    \midrule
    \textit{NVSB} & 0.000& 0.000 & \textbf{7.992} \\
    \textit{NVSB w/o $\mathcal{L}_{map2}$} &-0.060 & -0.090 & 8.040\\ %
    \bottomrule
    \end{tabular}

    \label{tab:ab2}
\end{table}

\subsubsection{Semi-supervised Learning}
To illustrate the advantage of the CVAE architecture that allows semi-supervised training, we compare NSVB trained with and without extra unpaired, unlabeled data on Chinese songs. The corresponding results are shown in Table \ref{tab:ab3}. The compared metrics indicate the advantage of semi-supervised learning, which facilitates the learning of the latent representations for better sample reconstruction (audio quality) and better latent conversion (vocal tone quality). %

\begin{table}[!htb]
    \small
    \centering
    \caption{The Comparison Mean Opinion Score in audio quality (CMOS-Q), vocal tone (CMOS-V) and the Mean Ceptral Distortion (MCD) of singing audio samples.}
    \begin{tabular}{ l | c | c | c}
    \toprule
    Method &  CMOS-Q & CMOS-V & MCD\\
    \midrule
    \textit{NVSB} & 0.000 & 0.000 & 7.068\\    
    \textit{NVSB w/o extra data} & -0.420 & -0.070  & 7.283\\  %

    \bottomrule
    \end{tabular}

    \label{tab:ab3}
\end{table}
\section{Conclusion}
In this work, we propose Neural Singing Voice Beautifier, the first generative model for the SVB task, which is based on a CVAE model allowing semi-supervised learning. For pitch correction, we propose a robust alignment algorithm: Shape-Aware Dynamic Time Warping (SADTW). For vocal tone improvement, we propose a latent mapping algorithm. To retain the vocal timbre during the vocal tone mapping, we also propose a new specialized SVB dataset named \datasetname~containing parallel singing recordings of both amateur and professional versions. The experiments conducted on the dataset of Chinese and English songs show that NSVB accomplishes the SVB task (pitch correction and vocal tone improvement), and extensional ablation studies demonstrate the effectiveness of the proposed methods mentioned above.

\bibliography{anthology}
\bibliographystyle{acl_natbib}

\appendix
\section{Details of Model Structure}
\label{appendxi:detailed_model}
The details of the adversarial discriminator, the content encoder, and WaveNet structure are shown in Figure \ref{fig:discriminator}, Figure \ref{fig:PPG}, and Figure \ref{fig:WN}.  The hidden size of CVAE model, latent variable and discriminator are 256, 128 and 128 respectively. We train NSVB on a single V100 32G GPU for almost 22 hours to finish two-stage training. 
\subsection{Multi-window Discriminator} 
\begin{figure}[htb]
    \centering
    \includegraphics[width=0.35\textwidth]{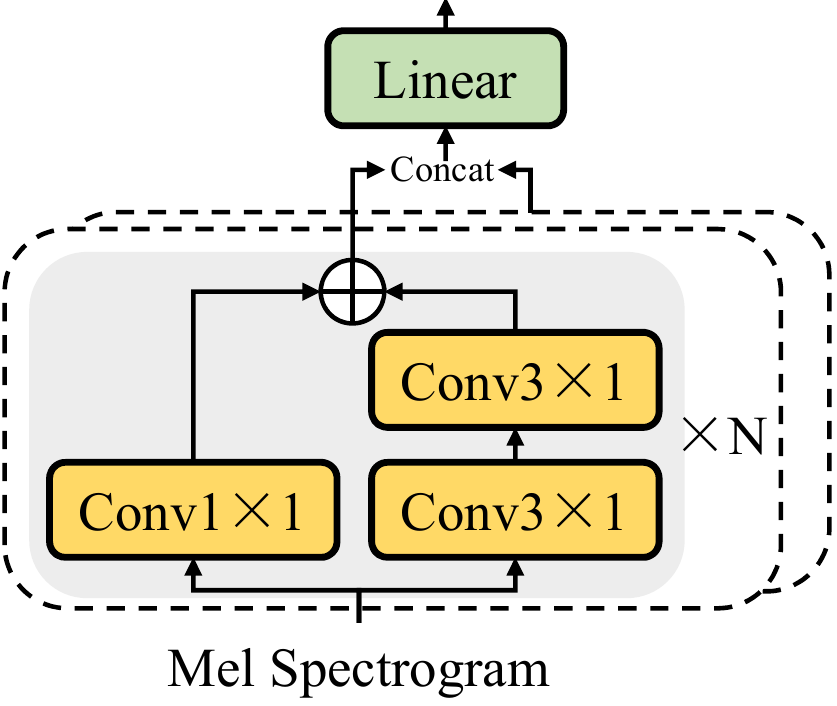}
    \caption{Multi-window discriminator structure used in NSVB}
    \label{fig:discriminator}
\end{figure}

As shown in Figure \ref{fig:discriminator}, our multi-window discriminator consists of 2 unconditional discriminator parts with fixed window sizes. Each unconditional discriminator contains $N$ layers of Conv units. In our model, we set $N=3$. The Conv units are all 1-D convolutional networks with ReLU activation and spectral normalization. The outputs of these unconditional discriminators are then concatenated and linearly projected to form the output.
\subsection{WaveNet}
\begin{figure}[!htb]
    \centering
    \includegraphics{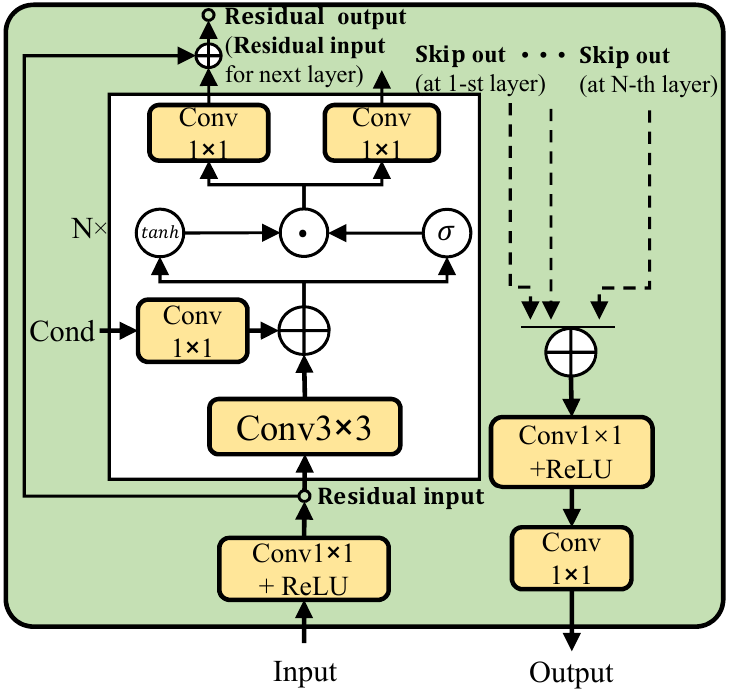}
    \caption{WaveNet structure used in NSVB}
    \label{fig:WN}
\end{figure}

As shown in Figure \ref{fig:WN}, the WaveNet unit used in the VAE encoder and decoder of NVSB consists of a 1D convolution layer with ReLU to preprocess the input, and a group of sub-layers with residual connection between adjacent ones. Each sub-layer contains a $1\times1$ convolutional layer to process the input condition and a $3\times3$ convolutional layer for residual input. After that, they got fused by being added up, then processed by tanh and sigmoid separately and then multiplied together. Finally, they produce a residual output for the next sub-layer and a skip-out. Lastly, two layers of 1D convolution and a ReLU process the summed skip-out to produce output. 

\begin{figure*}[!htb]
    \centering
    \begin{subfigure}{0.24\textwidth}
        \centering
        \includegraphics[width=1.0\textwidth]{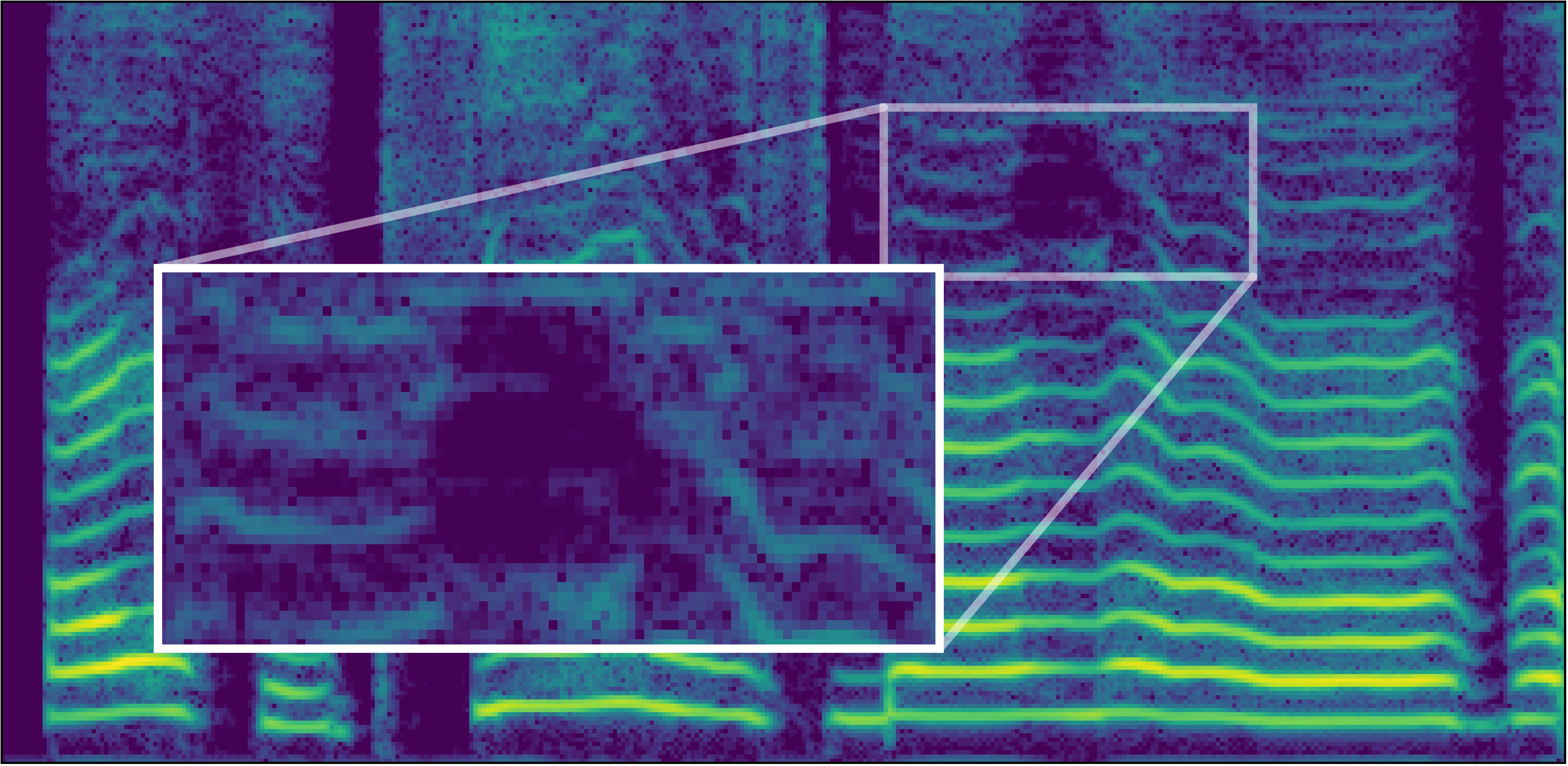}
        \caption{GT Amateur}
    \end{subfigure}
    \begin{subfigure}{0.24\textwidth}
        \centering
        \includegraphics[width=1.0\textwidth]{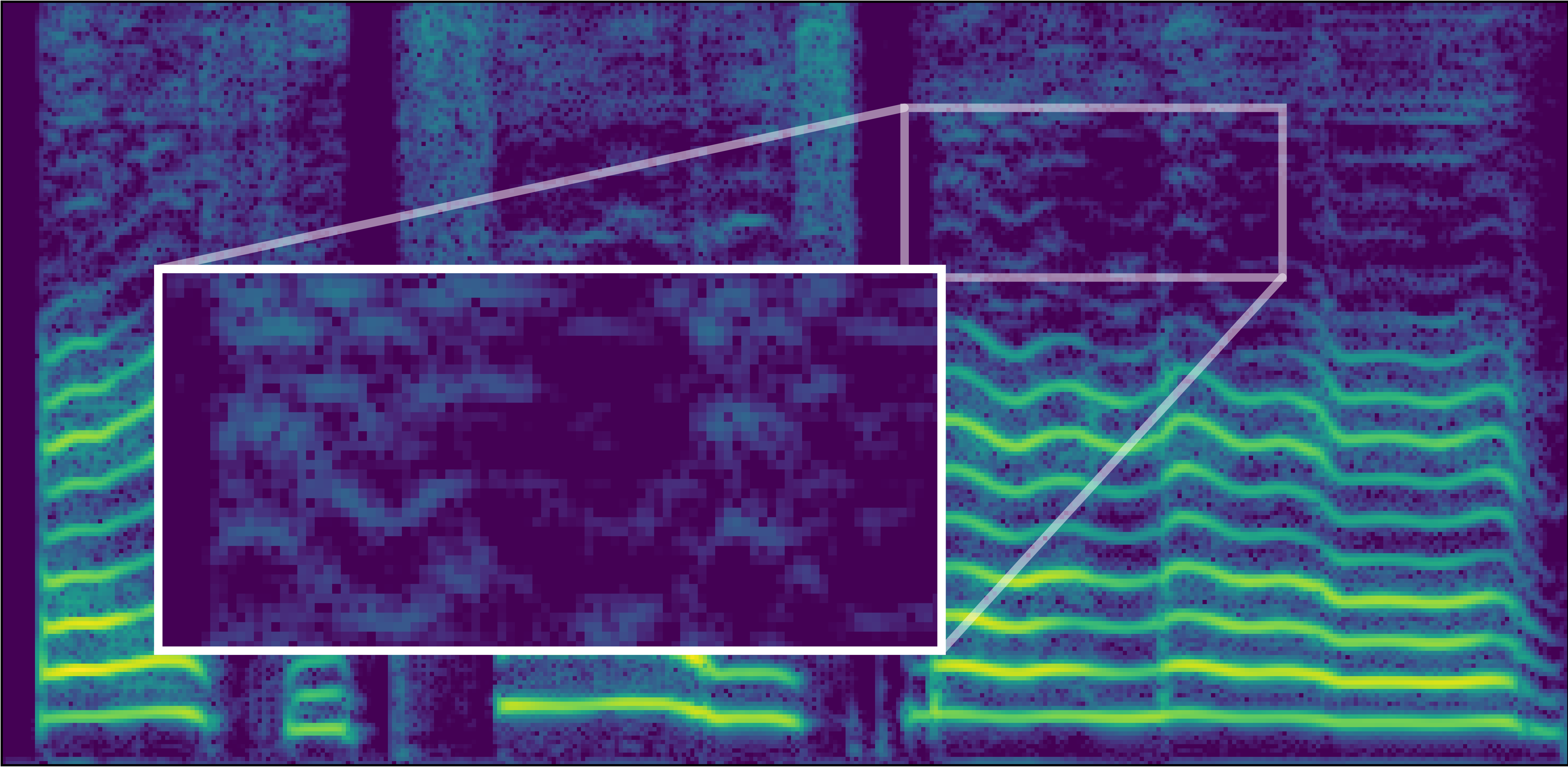}
        \caption{GT Professional}
    \end{subfigure}
    \begin{subfigure}{0.24\textwidth}
        \centering
        \includegraphics[width=1.0\textwidth]{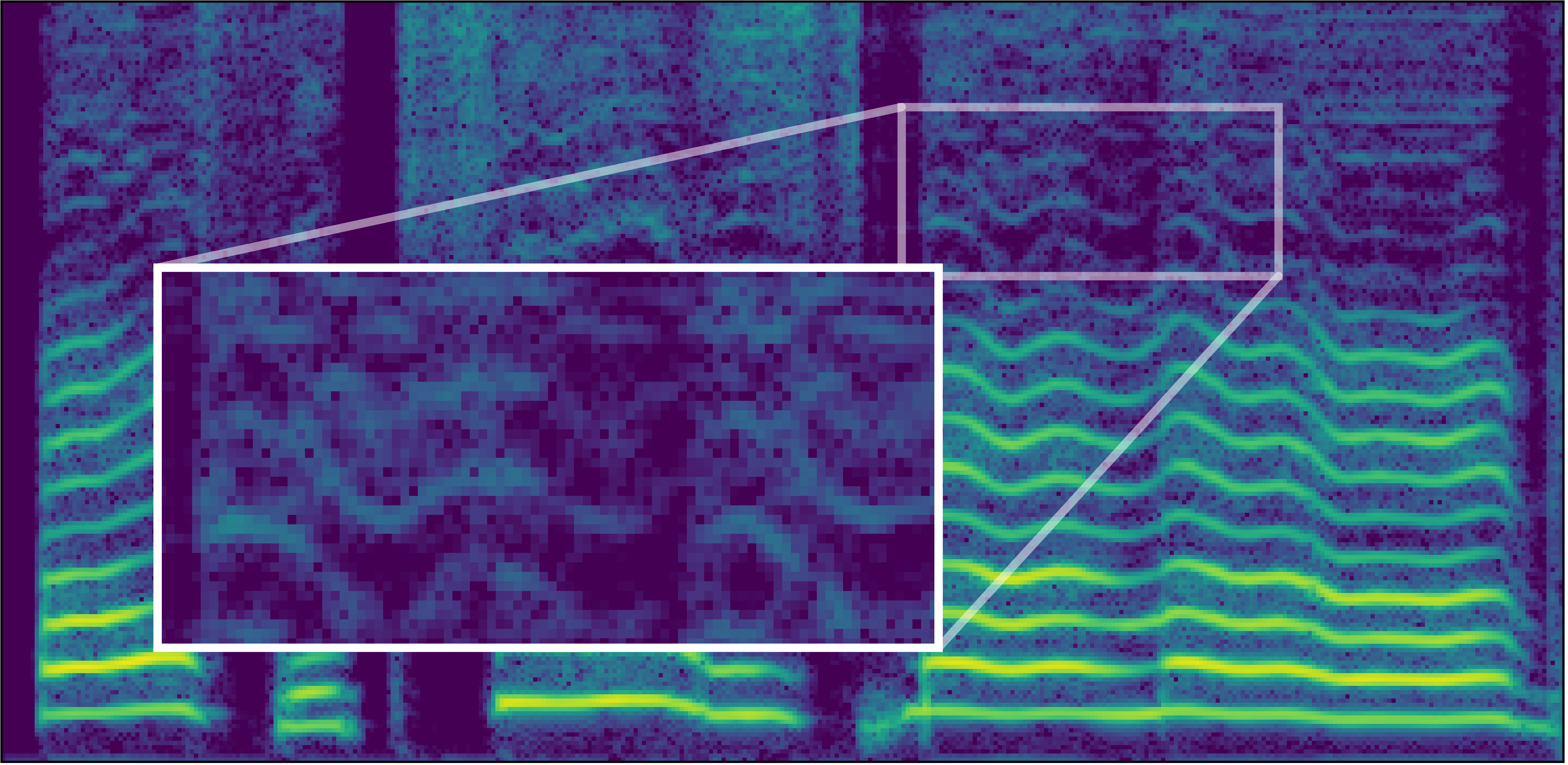}
        \caption{NVSB}
    \end{subfigure}
    \begin{subfigure}{0.24\textwidth}
        \centering
        \includegraphics[width=1.0\textwidth]{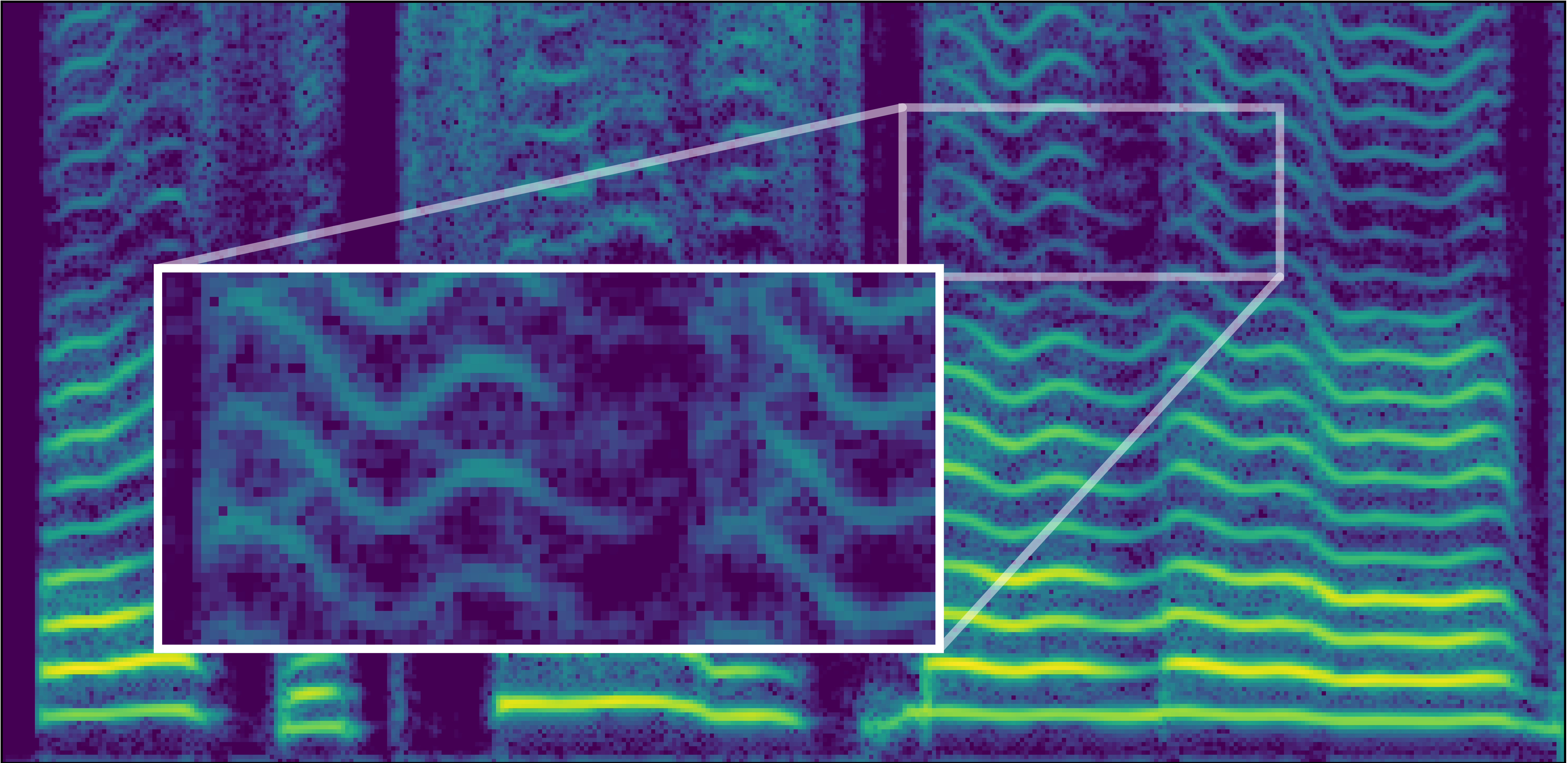}
        \caption{NVSB w/o mapping}
    \end{subfigure}
    \caption{Linear-spectrogram visualizations for the ablation study on latent mapping.}
    \label{fig:mels}
\end{figure*}

\subsection{Content Encoder}
\begin{figure}[htb]
    \centering
    \includegraphics[width=0.4\textwidth]{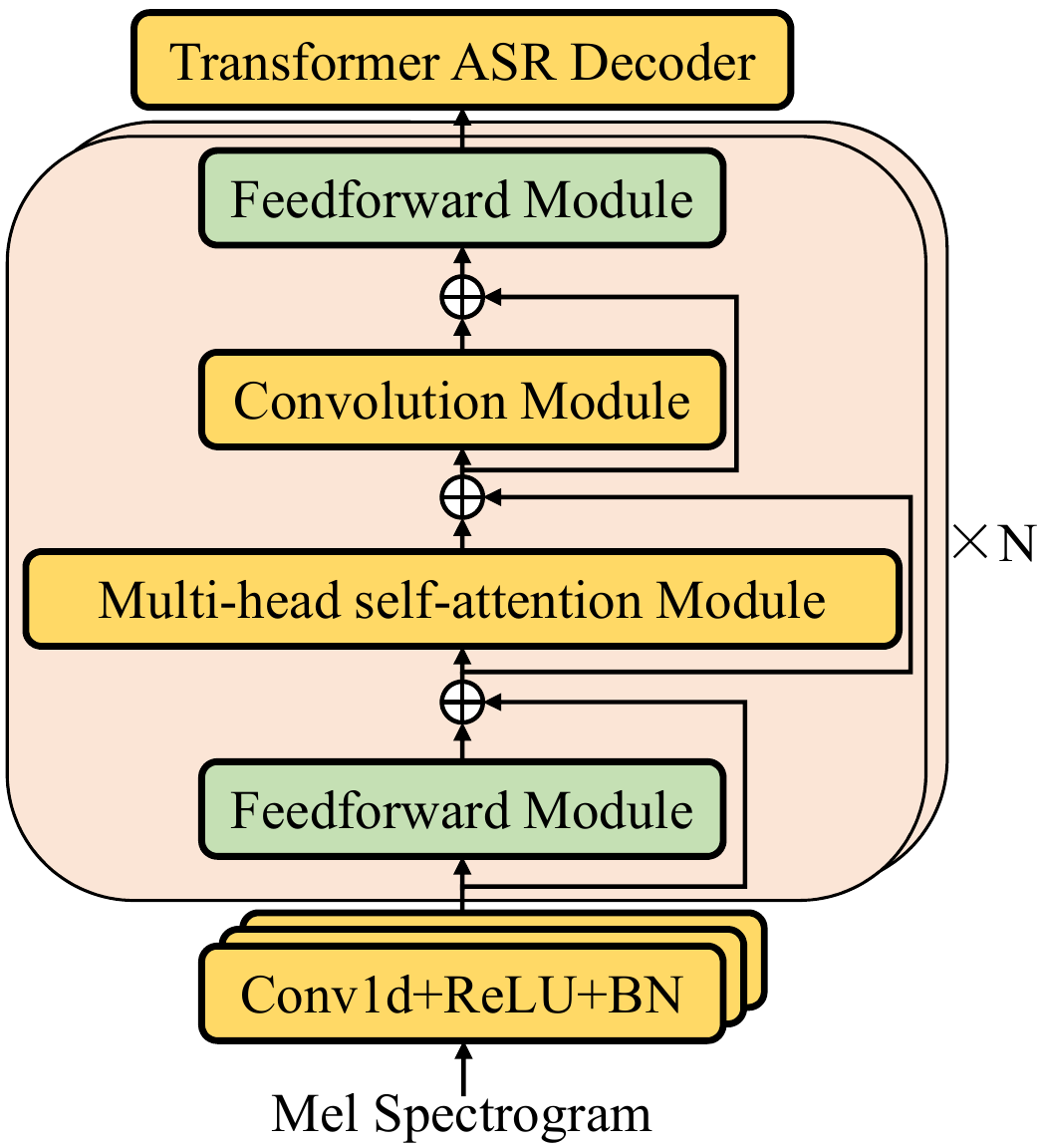}
    \caption{Content encoder used in NSVB}
    \label{fig:PPG}
\end{figure}

As shown in Figure \ref{fig:PPG}, the content encoder is the combination of several conformer encoder layers in pink rectangle along with a 3-layer prenet. The kernel size of the convolutional layer for prenet is 5. The hidden size is 256. We use 4 heads in the multi-head self-attention part. And we use 31 stacked conformer encoder layers to form this module. 
During pre-training, an ASR transformer decoder is attached to decode texts out for regular ASR training. After pre-training, only the encoder and the prenet part is used to extract PPG features from mel-spectrograms of audio samples.

\section{Linear-spectrograms Visualizations}
\label{appendxi:visualize}
We visualize four linear-spectrograms generated with the same content. It seems that the professional vocal tone is related to certain patterns in the high-frequency region of the spectrograms. In the future, SVB may be accomplished in a more fine-grained way together with the knowledge in vocal music.

\section{Details in subjective evaluations}
During testing, each audio sample is listened to by at least 10 qualified testers, all majoring in vocal music. We tell all testers to focus on one aspect and ignore the other aspect when scoring MOS/CMOS of each aspect.
For MOS, each tester is asked to evaluate the subjective naturalness of a sentence on a 1-5 Likert scale. For CMOS, listeners are asked to compare pairs of audio generated by systems A and B and indicate which of the two audio they prefer and choose one of the following scores: 0 indicating no difference, 1 indicating small difference, 2 indicating a large difference. For audio quality evaluation (MOS-Q and CMOS-Q), we tell listeners to "focus on examining the naturalness, pronunciation and sound quality, and ignore the differences of singing vocal tone". For vocal tone evaluations (MOS-V and CMOS-V), we tell listeners to "focus on examining singing vocal tone of the song, and ignore the differences of audio quality (e.g., environmental noise, timbre)". We split evaluations for main experiments and ablation studies into several groups for them.
They are asked to take a break for 15 minutes between each group of experiments to remain focused during subjective evaluations.
All testers get reasonably paid.
\label{appendxi:detailed_sub}

\end{document}